\documentclass[superscriptaddress,iop,apj]{emulateapj}
\usepackage{latexsym}
\usepackage{amssymb}
\usepackage{amsfonts}
\usepackage{threeparttable}                                                                                           
\usepackage{amsmath}
\usepackage{bm}
\usepackage[normalem]{ulem}
\usepackage{multirow}
\usepackage{natbib} 
\usepackage{subfigure}
\usepackage{color}
\usepackage{calc}
\usepackage{amsmath,amssymb,graphicx}
\usepackage{amssymb,amsmath}
\usepackage{tensor}
\usepackage{bm}
\usepackage{microtype}
\usepackage{booktabs}
\usepackage{times}
\usepackage[varg]{txfonts}
\usepackage[colorlinks, pdfborder={0 0 0}]{hyperref}
\usepackage{subfigure}
\usepackage{lipsum}
\usepackage{breakurl}
\usepackage{graphicx}
\usepackage{pbox}

\newcommand{\be}{\begin{equation}}
\newcommand{\ee}{\end{equation}}
\newcommand{\beq}{\begin{eqnarray}}
\newcommand{\eeq}{\end{eqnarray}}

\newcommand{\msun}{\,M_{\odot}}

\begin{document}

\title{Further Evidence for a Weak Neutron Star Magnetosphere in the Accreting Millisecond X-Ray Pulsar HETE J1900.1-2455}

\shortauthors{Patruno et~al.}

\author{
  A. Patruno\altaffilmark{1,2} \& R. Wijnands\altaffilmark{3}}
  \altaffiltext{1}{Leiden Observatory, Leiden University,
              Neils Bohrweg 2, 2333 CA, Leiden, The Netherlands}
  \altaffiltext{2}{ASTRON, the Netherlands Institute for Radio Astronomy, Postbu 2, 7900 AA, Dwingeloo, the Netherlands}
  \altaffiltext{3}{Anton Pannekoek Institute for Astronomy, University of Amsterdam, Postbus 94249, 1090 GE Amsterdam, The Netherlands}

\title{Further Evidence for a Weak Neutron Star Magnetosphere in the Accreting Millisecond X-Ray Pulsar HETE J1900.1-2455}

\begin{abstract}
\noindent
HETE J1900.1--2455 is a peculiar accreting millisecond X-ray pulsar
(AMXP) because it has shown intermittent pulsations after 22 days from
the beginning of its outburst. The origin of intermittent pulses in
accreting systems remains to be understood. To better investigate the
phenomenon of intermittent pulsations here we present an analysis of 7
years of X-ray data collected with the \textit{Rossi X-Ray Timing
  Explorer} and focus on the aperiodic variability. We show that the
power spectral components follow the same frequency correlations as
the non-pulsating atoll sources. We also study the known kHz QPO and
we show that it reaches a frequency of up to $\approx900$ Hz, which is the
highest frequency observed for any kHz QPO in an AMXP. We also report
the discovery of a new kHz QPO at $\approx500$ Hz. Finally, we discuss
in further detail the known pulse phase drift observed in this source,
which so far has no explanation. We interpret the behavior of the
aperiodic variability, the high frequency of the 900 kHz QPO and the
presence of the pulse drift as three independent pieces of evidence
for a very weak neutron star magnetosphere in HETE J1900.1--2455.

\end{abstract}
\keywords{pulsars: individual HETE J1900.1-2455 -- X-Rays: binaries -- X-Rays: accretion} 

\maketitle

\section{Introduction}

The accreting millisecond X-ray pulsar (AMXP) HETE J1900.1--2455
(henceforth refereed to as J1900) is a binary with an orbital period
of 1.4 hr composed by a neutron star spinning with a frequency of
about 377 Hz and a 0.02 $M_{\odot}$ brown dwarf.  This source is a
very peculiar one among the AMXP family (see e.g.,~\citealt{pat12r}
for a review on AMXPs). J1900 is a quasi-persistent source (i.e.,
showing years-long outbursts) whose first outburst was detected on
June 14, 2005 by the {{\it High Energy Transient Explorer-2}}
\citep{vand05} and its outburst has lasted for approximately 10 years
\citep{deg17}.  Two days after its discovery, observations taken with
the \textit{Rossi X-ray Timing Explorer} ({{\it RXTE}}) showed X-ray
pulsations~\citep{kaa06}. The pulsations have a unique and peculiar
behavior as they were first continuously detected for approximately 22
days \citep{kaa06}, then they became intermittent \citep{gal07, gal08}
appearing and disappearing on a timescale of a few days. After a few
tens of days the pulsations became extremely intermittent, appearing
sporadically at very low amplitudes for the next 2
years~\citep{pat12c} and finally they became undetectable with upper
limits of $\lesssim0.5-0.1$\% rms on the pulsed amplitudes for the
remaining time until December 5th, 2011, when {\it RXTE} stopped
monitoring the source.  A simultaneous \textit{RXTE} and
\textit{XMM-Newton} observation performed on September 19--20, 2011 also
revealed no pulsations down to 0.4\% amplitude~\citep{pap13d}.

\citet{kaa06} reported the appearance of a 882 Hz
Quasi-Periodic-Oscillation (the so-called kHz QPO) in coincidence with
a relatively large flare in the X-ray lightcurve of J1900, occurring
about 22 days after the beginning of the monitoring campaign. As noted
by \citet{gal07}, pulse intermittency sets in right after this flaring
episode. \citet{pat12c} also found a very large pulse phase shift
($\simeq-0.7$ cycles) in coincidence with the flare.  The pulse phases
drifted roughly constantly for $\approx3000$ seconds, until they
reached a phase of $\phi\approx-0.1$ cycles towards the end of the
observation. Whether the flare, the kHz QPO and the pulse phase shift
are all related to each other is currently unknown.  It is important
to highlight that pulsations are seen in J1900 only in the
island/extreme-island (hard) state\footnote{Atoll sources trace out a
  pattern in the color-color diagram with a curved branch called the
  banana branch, further divided in the upper and lower banana branch (at
  the highest intensities) which, at the lowest intensity, connects to the island and extreme island state.~\citep{has89,van03,van05}.} whereas they
disappear in the banana branch (soft-state) except during the
aforementioned flare that marked the onset of
intermittency~\citep{gal07, alt08b}. This could indicate that there is a link
between the presence of pulsations and the spectral state of the
source.

Another peculiar behavior of J1900 is that the accretion torques
operating on the neutron star decrease exponentially with
time~\citep{pat12c}. This has been interpreted as evidence for
magnetic field burial due to accretion (see \citealt{cum01}) that
could be responsible for reducing the strength of the neutron star
magnetosphere. Intermittency and the decreasing strength of the
magnetic field might also be related to the peculiar behavior of the
burst oscillations seen in J1900.  The source has shown several tens
of thermonuclear X-ray bursts but only in one case burst oscillations
have appeared~\citep{wat09b}. Such burst oscillations were detected on
April 2, 2009, well beyond the point were the last pulsations were
seen.  The burst oscillations show the typical drift in their
frequency of about 1 Hz, similar to that seen in other bursting
non-pulsating X-ray binaries and markedly different than burst
oscillations in other AMXPs~(see~\citealt{gal08b,wat12} for reviews). An
interesting point here is that the burst oscillations appeared during
a burst that occurred during another flare, similar to the one
described before, when the source moved to the banana branch.
Pulsations were not detected around the time of the burst.

A poorly explored observational diagnostic in HETE J1900.1--2455, that
could help to solve some of these puzzles, is the aperiodic
variability of the X-ray lightcurve.  Indeed it is known that kHz QPOs
and other power spectral frequency components are powerful probes of
the accretion flow close to the compact object (where magnetospheric
effects could be present).  For example, the characteristic
frequencies of some of the power spectral components of AMXPs (in
particular the highest frequency components) are shifted with
respect to other non-pulsating atoll sources by a factor 1.1-1.6,
depending on the specific source~\citep{van03, van05, lin05,bul15}.
It is possible that it is the magnetic field of the neutron star
that plays a role in generating these shift and therefore 
it is interesting to investigate this phenomenon in J1900 too. 

In this paper we explore all the {{\it RXTE}} data collected
on J1900 between June 2005 and December 2011 to look for new
signatures that might highlight if the magnetosphere in J1900 is
typical as in other AMXPs.  In particular we look for kHz QPOs, the
frequency correlations between different power spectral components and
we link the observations to the behavior of the pulsations.

\section{X-Ray Observations}\label{sec:obs}

We used all the \textit{RXTE} Proportional Counter Array (PCA) data
recorded over a baseline of 6 years (2005 June to 2011 December).  We
refer to \citet{jah06} for PCA characteristics and \textit{RXTE}
absolute timing. For our aperiodic timing analysis we used Event and
GoodXenon data, with a resolution of $2^{-13}$ s and $2^{-20}$ s,
respectively. We rebinned the GoodXenon data to the same resolution as
the Event data. For the timing analysis, the photons were extracted
from the energy band $\approx2$--$16$ keV (absolute channels 5--37)
which is the one that maximizes the signal to noise (S/N) ratio of the
pulsations ~\citep{pat12c} and, as we verified a-posteriori, gives the
optimal S/N also for the aperiodic variability.  Power spectra were
calculated selecting intervals with lengths of 16 s (for the dynamical
power spectra) and 128 s (for all other power spectra).

No background is subtracted and no dead time correction is applied
prior to calculating the power spectra.  After their calculation, we
average and Leahy-normalize them by subtracting a Poissonian
(counting) noise level that incorporates dead-time effects as
explained in \citet{zha95}.  We then rms-renormalize the power by
following \citet{van95} and using an average background count-rate
calculated by processing Standard-2 data with the FTOOL
\emph{pcabackest} . In essence, the final power spectra are such that
the square root of the integrated power equals the fractional
root-mean-square variability in the signal.  The power spectra are
averaged per observation (obsID) and inspected by eye and those that
show significant power are fitted by using a sum of Lorentzians, whose
peak frequency is defined according to the definition
of~\citet{bell02}:
\begin{equation}
\nu_{max}=\nu_0\sqrt{+\frac{1}{4Q^2}}
\end{equation}
 where $\nu_0$ is the centroid frequency and $Q$ is the quality
 factor.  This latter is defined as the ratio between $\nu_0$ and the
 full-width at half maximum of the QPO.  In this paper we define the
 single trial significance $\sigma$ of a QPO by dividing the integral
 power of the QPO with its negative error, as is common practice in
 aperiodic timing studies. The errors on the integral power are
 determined by varying the minimum $\chi^2$ such that
 $\Delta\chi^2=1$.

To increase the signal-to-noise (S/N) ratio of the broader power
spectral features and investigate correlations between them, we then group
the observations in such a way that each group contains consecutive
observations close in time and in color. When grouping, we allow for a
variation of a few percent in color, to guarantee that no spectral
transition occurs in each group. Since J1900 has been sparsely
monitored, with gaps in consecutive data reaching several days/tens of
days, sometimes we use single obsIDs (that we still label as
``groups'' to keep a uniform nomenclature, see group 3 and 8 in
Table~\ref{tab:obs}).  We also tried to select groups by choosing
observations close in soft and hard colors, regardless of their
observing time. However, this grouping scheme does not lead to a
better S/N of the different power spectral components, which are often
blurred together. This leaves us with a total of 62 groups. 

 We use the Standard 2 mode data (16 s time resolution) to calculate
 the aforementioned X-ray colors and intensity. Hard and soft colors
 are defined, respectively, as the 9.7--16.0 keV/6.0--9.7 keV and the
 3.5--6.0 keV/2.0--3.5 keV count rate ratios, whereas the intensity is
 extracted in the 2--16 keV band. The energy/channel conversion is
 done by using the pca\_e2c\_e05v02 table provided by the {{\it RXTE}}
 team. The colors and intensity are calculated after subtracting the
 background and correcting for dead-time and are normalized to the
 Crab Nebula values to account for gain changes among the different
 PCUs that compose the PCA instrument (see \citealt{kuu94};
 \citealt{van03}).
\begin{figure}
  \centering
  \rotatebox{-90}{\includegraphics[width=0.88\columnwidth]{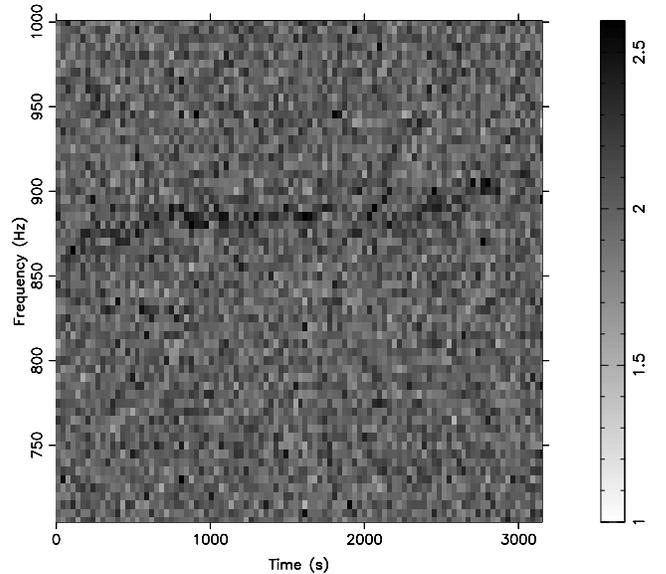}}
  \caption{Dynamical power spectrum of HETE J1900.1-2455 for the
  ObsId 91015-01-06-00. This is the observations where the first flare is seen and when the pulses are observed to drift by about 0.5 cycles in $\approx 3000$ s. At the same time the kHz QPO moves from $\sim850$ to $\sim900$ Hz.}\label{fig:dyn}
\end{figure}

Finally, when considering the pulse phase drift that occurred at the
time of the first flare, we generate the pulsations by following the
procedure explained in \citet{pat12c}. In short, we fold the photon
time of arrivals with the orbital and spin frequency parameters of
J1900. The only difference here is that we introduce an energy
selection for the photons to study the pulse phase energy dependence.
We define a soft and a hard band, with absolute channels 5--23
($\approx2$--$9$ keV) and 24--39 ($\approx9$--$17$ keV), respectively. To
partially compensate for the loss in S/N due to the energy band
selection, we fold the data in segments of 500 s instead of the 300 s
chosen by \citet{pat12c}. The pulsations are considered significant if
the ratio between their amplitude and their statistical error is
larger than 3$\sigma$.

\section{Results}

\subsection{Aperiodic Variability}

\begin{figure*}
  \centering
  \rotatebox{-90}{\includegraphics[width=0.72\columnwidth]{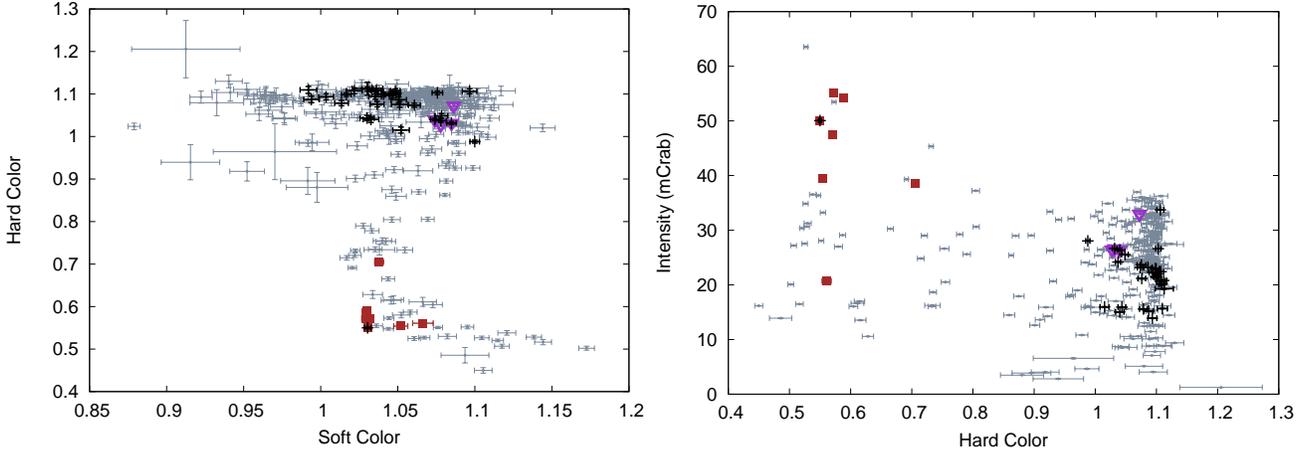}}
  \caption{Color-color diagram (left panel) and hardness-intensity diagram (right panel) for all the data available for HETE J1900.1--2455 averaged per ObsID and selected such that the intensity is larger than at least 1 mCrab. The red squares are points where the 800--900 kHz QPO is detected. The purple open triangles identify the observations where the 400--500 kHz QPO is observed. The black circles are points where pulsations are seen.}
  \label{fig:cc}%
\end{figure*}

We begin by discussing the presence of kHz QPOs in our power spectra, 
with the results of our analysis presented per observation (obsID).
We do this because we want to see if there is a link between the presence
of kHz QPOs and the presence of pulsations. Since pulsations appear and
disappear on a timescales as short as few minutes, we cannot use
the grouped data (which will be used later to look for correlations
among the different power spectral components). 
We detect several kHz QPOs in different observations that can be
clearly classified in two categories. The first comprises six
observations in which one single kHz QPO is detected (including the
882 Hz QPO previously reported by \citealt{kaa06}) with a peak
frequency always between $\approx$780 and 900 Hz.  The QPO frequency
$\nu_{max}$ is seen to quickly drift by several tens of Hz during some
observations (see for example the dynamical power spectrum in
Figure~\ref{fig:dyn}).  This QPO is seen in the lower left banana
state (i.e., the soft state, see Figure~\ref{fig:cc}) and it is reminiscent of other kHz QPOs
seen in atoll sources \citep{van06}. The kHz QPOs are detected with a
significance (single trial) between 3.1 and 8$\sigma$. No evidence for
any other kHz QPO with frequencies between 100 Hz and the Nyquist
frequency is seen in these six observations.  We also detect a 883 Hz
QPO seen in the island state (with a significance of 4.1$\sigma$) in
the obsID 92049-01-44-00.

\begin{figure*}[ht!]
  \begin{tabular}{cc}
    \includegraphics[width=0.5\textwidth]{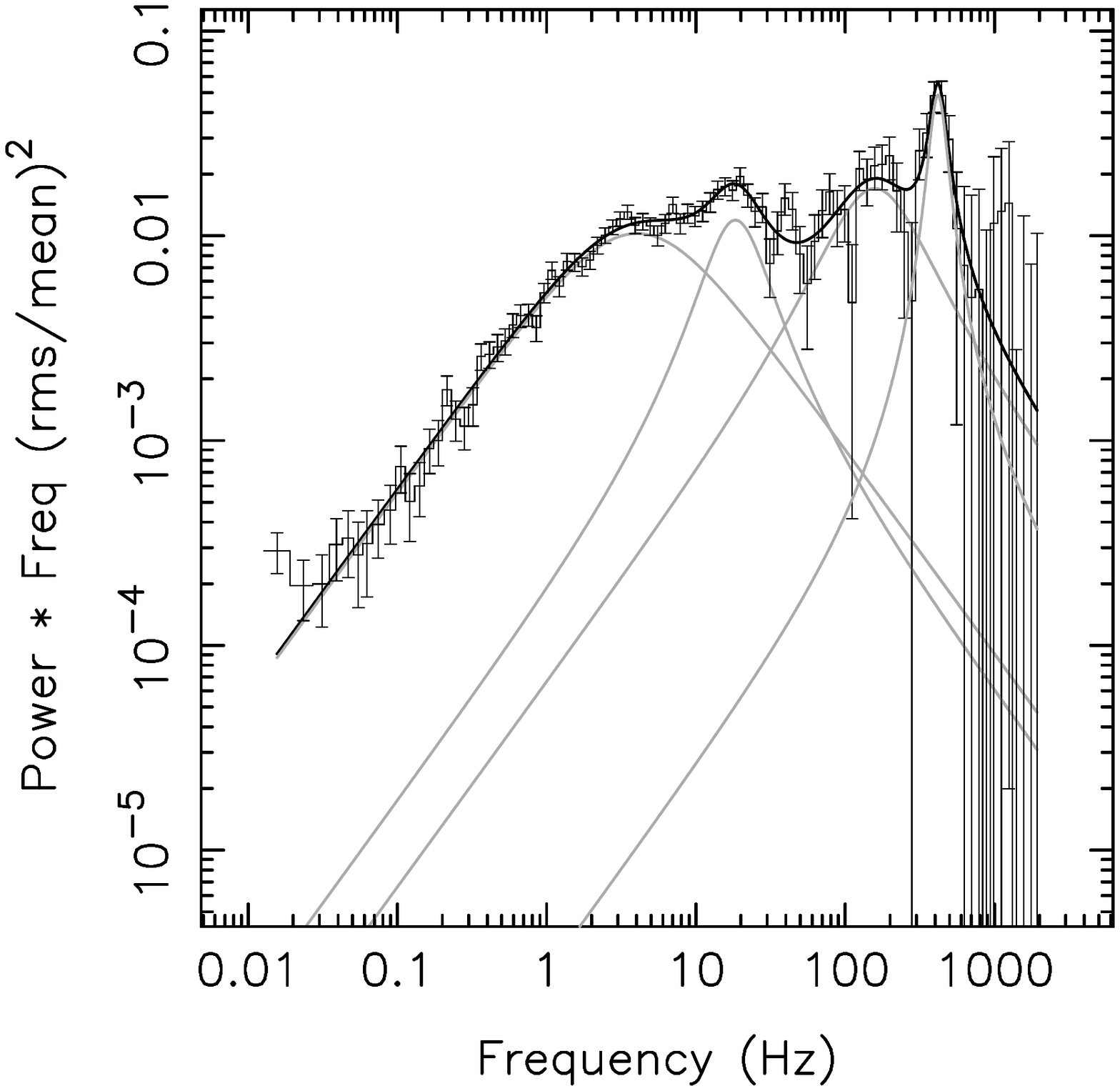}& \includegraphics[width=0.485\textwidth]{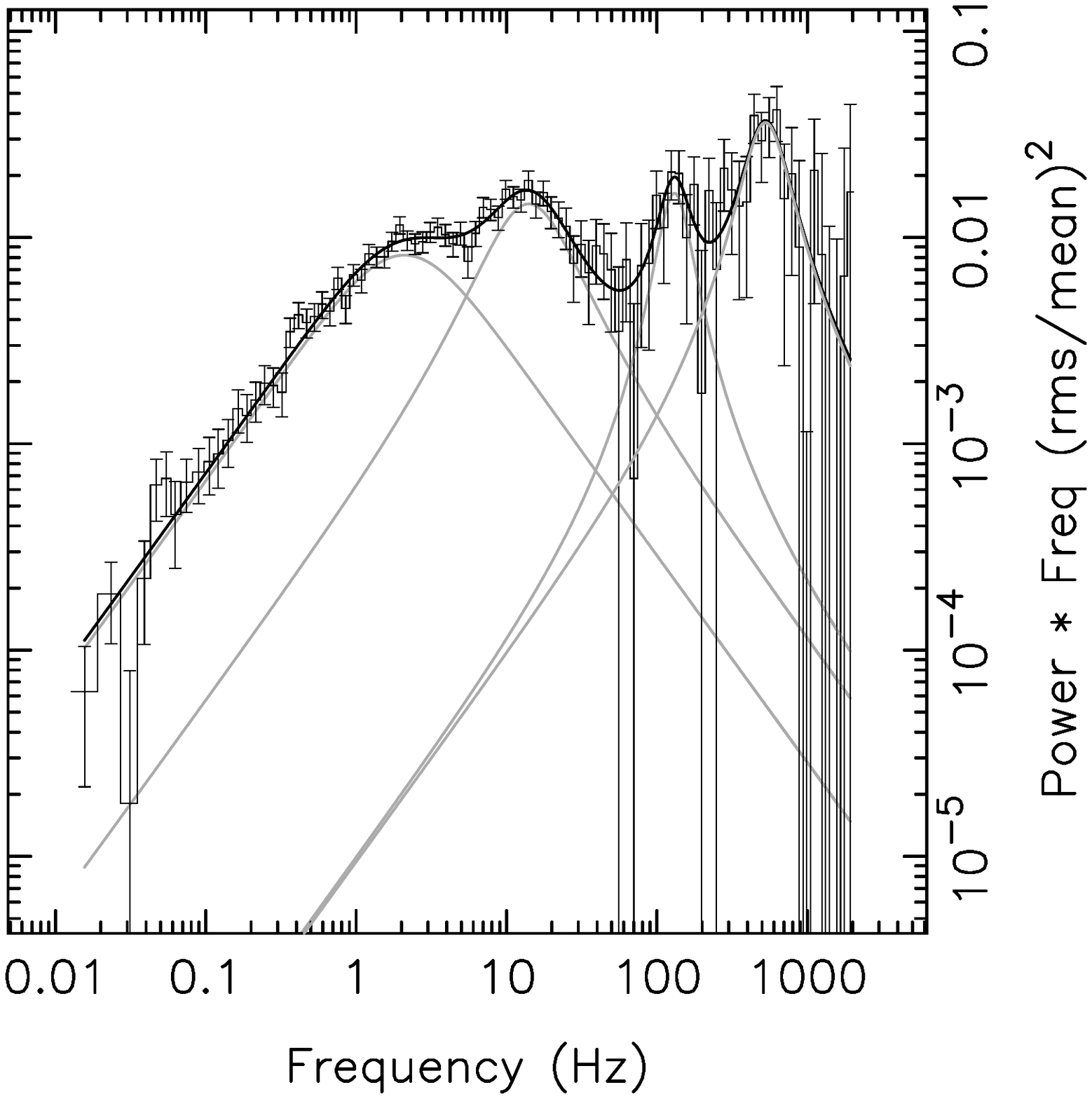}\\
  \end{tabular}
  \caption{ Power density spectra of the observations 91015-01-04-04 (left) and 91057-01-04-00 (right). In the plots it's evident 
the new detected kHz QPO. The frequency of the QPO centroid shifted from $\sim 420$Hz (left panel) to $\sim 510$Hz (right panel) with the first flare occurring in between these two observations.}\label{fig:pds}
\end{figure*}

Another observation taken in the island state (ObsID 91015-01-06-02)
and recorded four days since the onset of intermittency, shows also a
possible kHz QPOs with frequency of $\approx720$ Hz, although the
significance of the detection is low (3.2$\sigma$). No pulsations
were detected in this observation. Given the low significance of the
feature we consider the QPO detection as tentative and will not
discuss it any further.
The second category of QPOs is composed by a kHz QPO with a frequency of
$\approx400-500$ Hz which appears sometimes in pair with a low
frequency (and low Q) feature at around 100--200 Hz during the 
island state (see an example in Figure~\ref{fig:pds}).

\begin{figure*}
  \centering
  \rotatebox{-90}{\includegraphics[width=1.5\columnwidth]{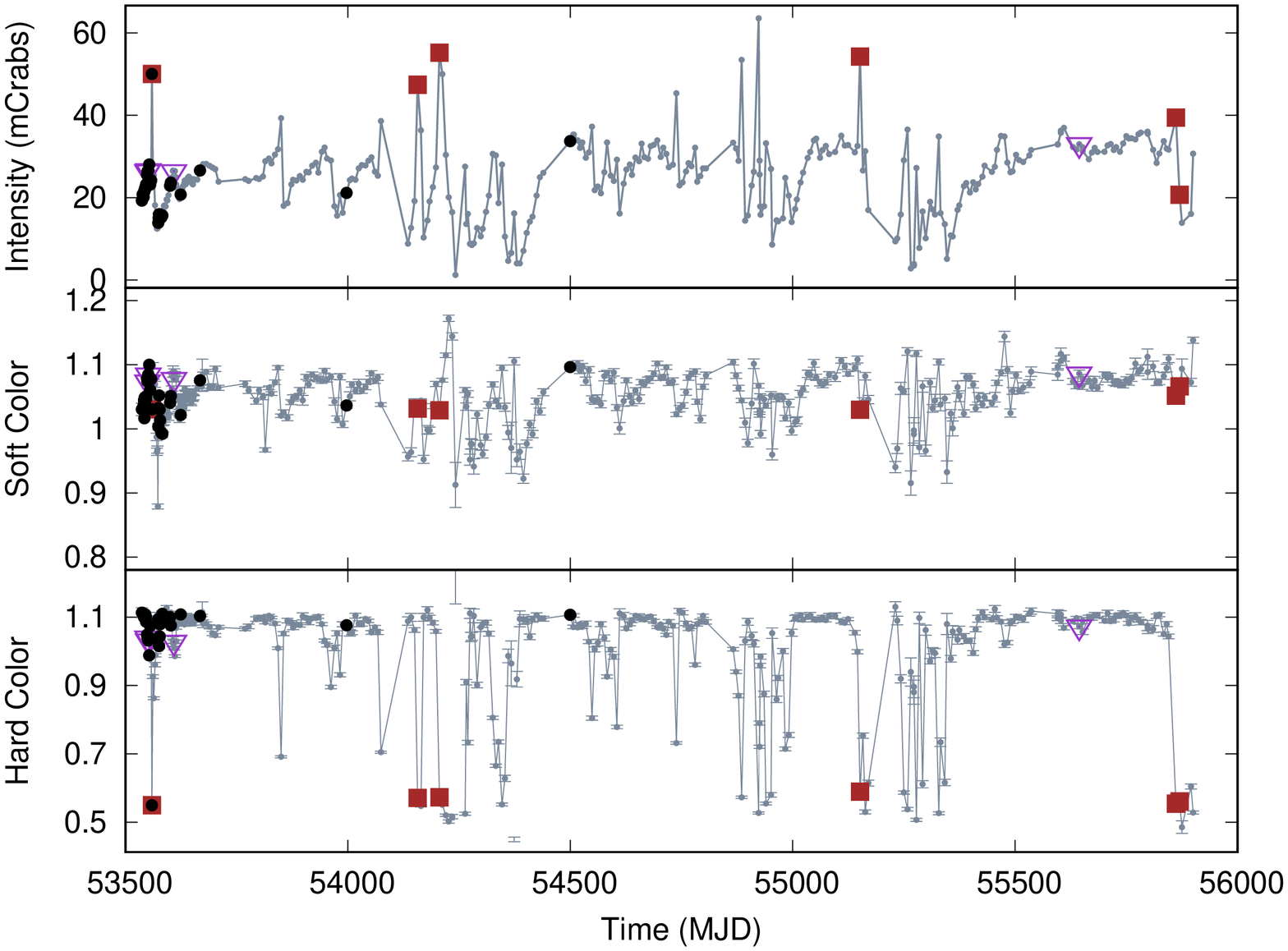}}
  \caption{The 2--16 keV lightcurve (top panel), soft color (middle
    panel) and hard color (bottom panel) for all observations recorded
    by \textit{RXTE} from 2005 to 2011. Observations with pulsations
    are marked with black circles, the ones with the 800-900 kHz QPO
    with red squares and those with the 400-500 kHz QPO are identified
    by open purple triangles.}
  \label{fig:3pl}%
\end{figure*}

\begin{figure}
  \centering
  \rotatebox{-90}{\includegraphics[width=0.7\columnwidth]{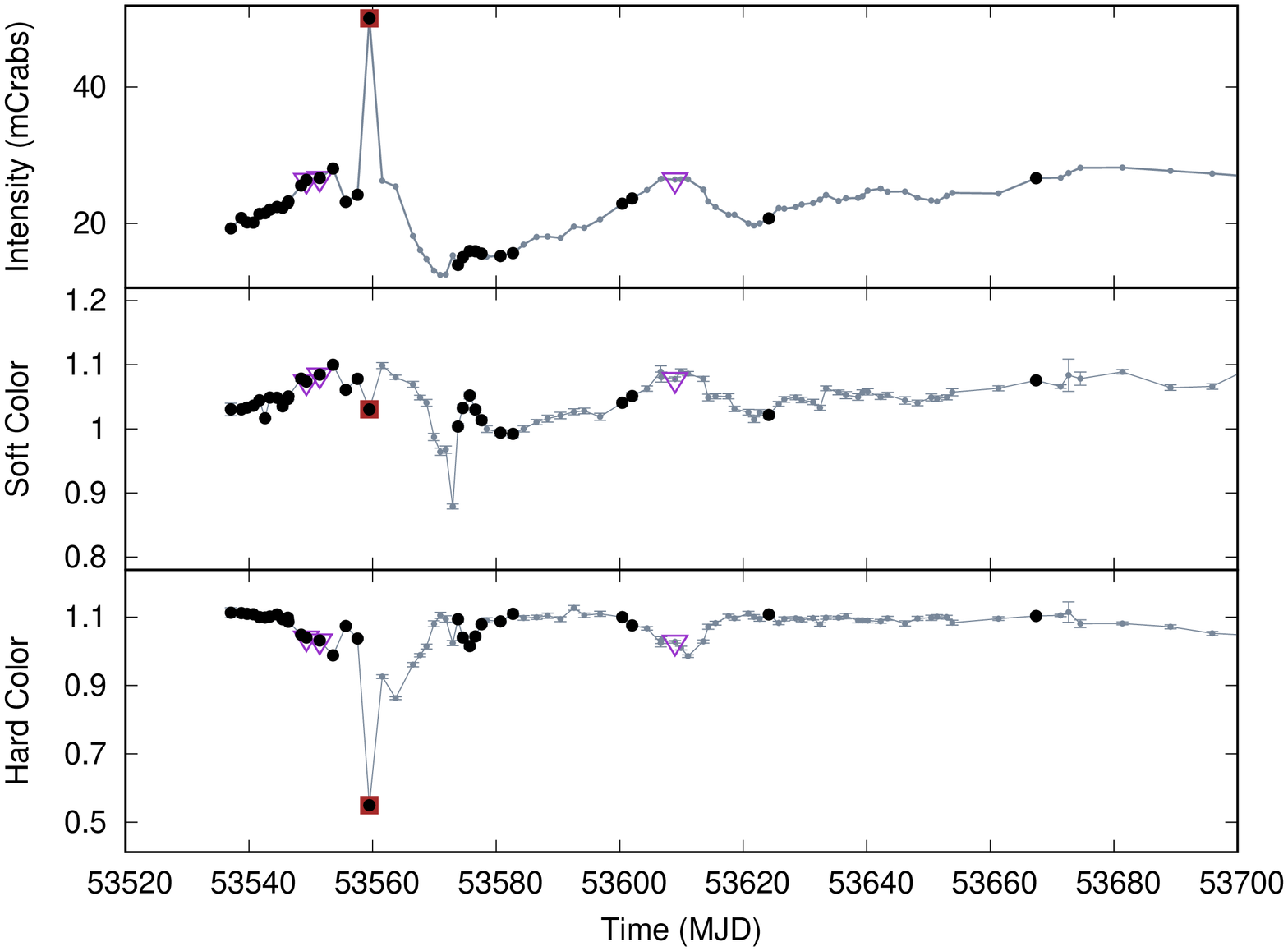}}
  \caption{A zoom in of the lightcurve in Figure~\ref{fig:3pl} to better display the initial part of the
  outburst.}
  \label{fig:3plzoom}%
\end{figure}

If these are indeed twin kHz QPOs pairs, then the difference
$\Delta\nu$ between the lower and upper kHz QPO frequency varies
between 260 and 380 Hz.  However, atoll sources very often show a
feature around 100--200 Hz which is called the hecto-Hertz QPO, and we
consider it more likely that this is what we are observing in J1900
(see further discussion below). The hHz QPO has the characteristic of
having a frequency relatively constant and to be uncorrelated with the
other characteristic frequencies seen in the power spectra (see e.g.,
\citealt{van05}). In Figure~\ref{fig:3pl} and \ref{fig:3plzoom} we
show the lightcurve of J1900 along with the observations were the
different QPOs were detected.

The color-color and hardness-intensity diagram of J1900 (shown in
Figure~\ref{fig:cc}) exhibit a typical atoll source
behavior, with the source that spends most of its time in the extreme
island state and banana branch at higher luminosity.  Based on our
previous analysis reported in \citet{pat12c}, we highlighted all
observations where pulsations were detected.  We also highlight
observations where the 800--900 Hz QPO is detected (red squared) plus
those observations with the 400--500 Hz QPO (purple triangles).


We now focus on all the different power-spectral components observed
in J1900. Since these components are usually broad we grouped the data
as explained in Section~\ref{sec:obs} to increase the S/N ratio.  To
better investigate the origin of the kHz QPOs as well as the different
power spectral components we fit a multi-Lorentzian empirical model to
determine the temporal evolution of the different characteristic
frequencies. We name the different power spectral components by
following the work of \citet{van03,van05}. In brief, power spectra are
usually composed by the following Lorentzians ($L$): a low-frequency
break ($L_b$), a second break that often appears at high luminosity
($L_{b2}$), the hump ($L_h$), the hecto-Hz QPO ($L_{hHz}$), the
low-frequency QPO ($L_{\rm LF}$), a low frequency component
($L_{low}$) which is a possible manifestation of the lower kHz QPO,
the lower kHz QPO ($L_l$) and the upper kHz QPO ($L_u$).  We
anticipate that we do not see all these features in J1900 but we need
usually 3 to 4 Lorentzians to obtain a good fit to our power-spectra.
In particular, we do not see $L_{\rm LF}$ and $L_{l}$.

\begin{figure}
  \centering
  \rotatebox{-90}{\includegraphics[width=0.7\columnwidth]{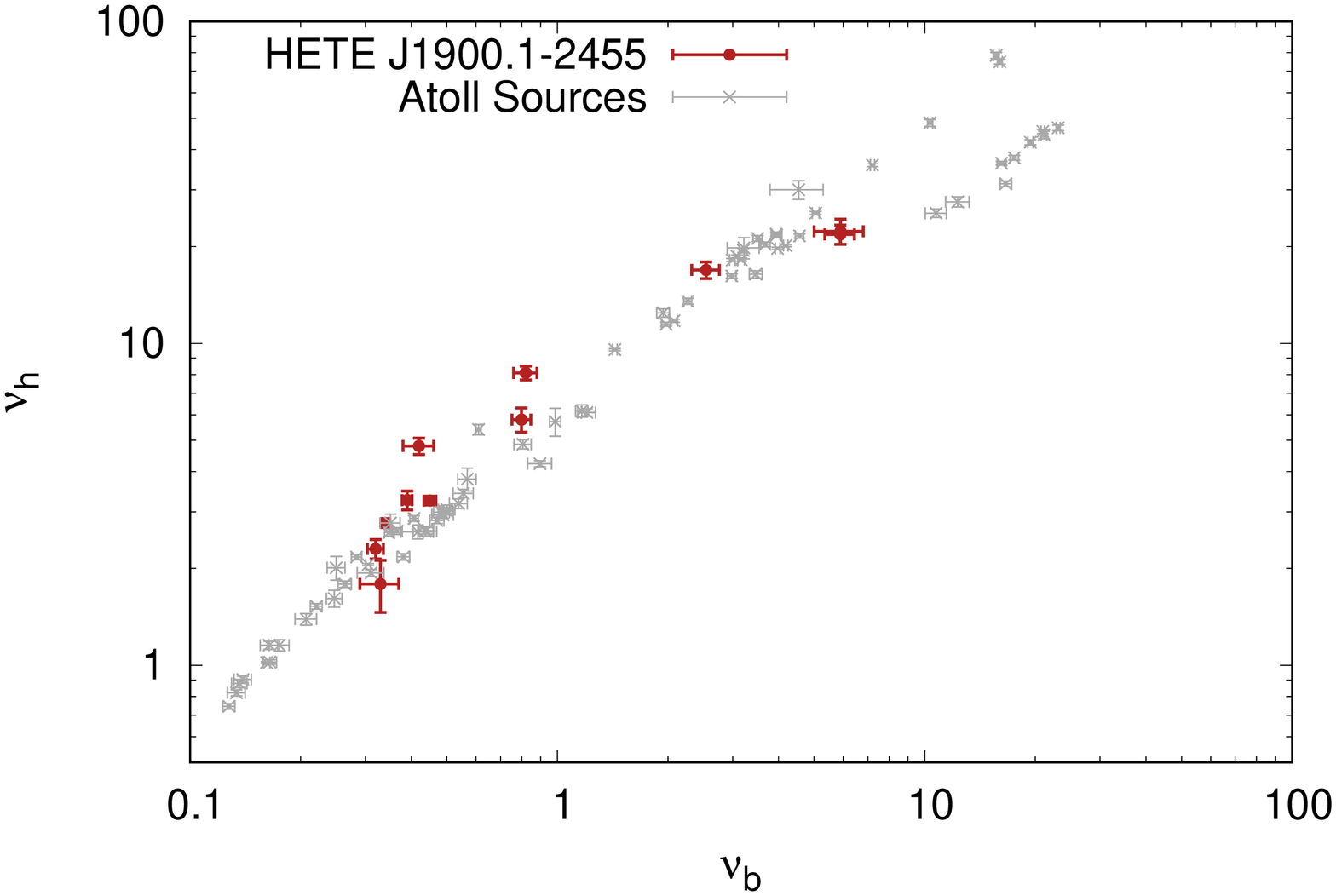}}
  \caption{WK relation for HETE J1900.1-2455 (red points) over-plotted on the other Atoll sources (gray points, from \citealt{van05}).}
  \label{fig:wk}%
\end{figure}

We begin by identifying the low frequency components in each power
spectrum, namely $L_b$, $L_{b2}$ and $L_h$. The break and the hump
characteristic frequencies are indeed known to follow the so-called WK
correlation~\citep{wij99}.  These frequencies measured in J1900 match
with great accuracy the known correlation for other atoll sources (see
Figure~\ref{fig:wk}). This means that our identification of those
components in the power spectra of J1900 is correct.

\begin{figure*}
  \centering
  \rotatebox{-90}{\includegraphics[width=1.0\columnwidth]{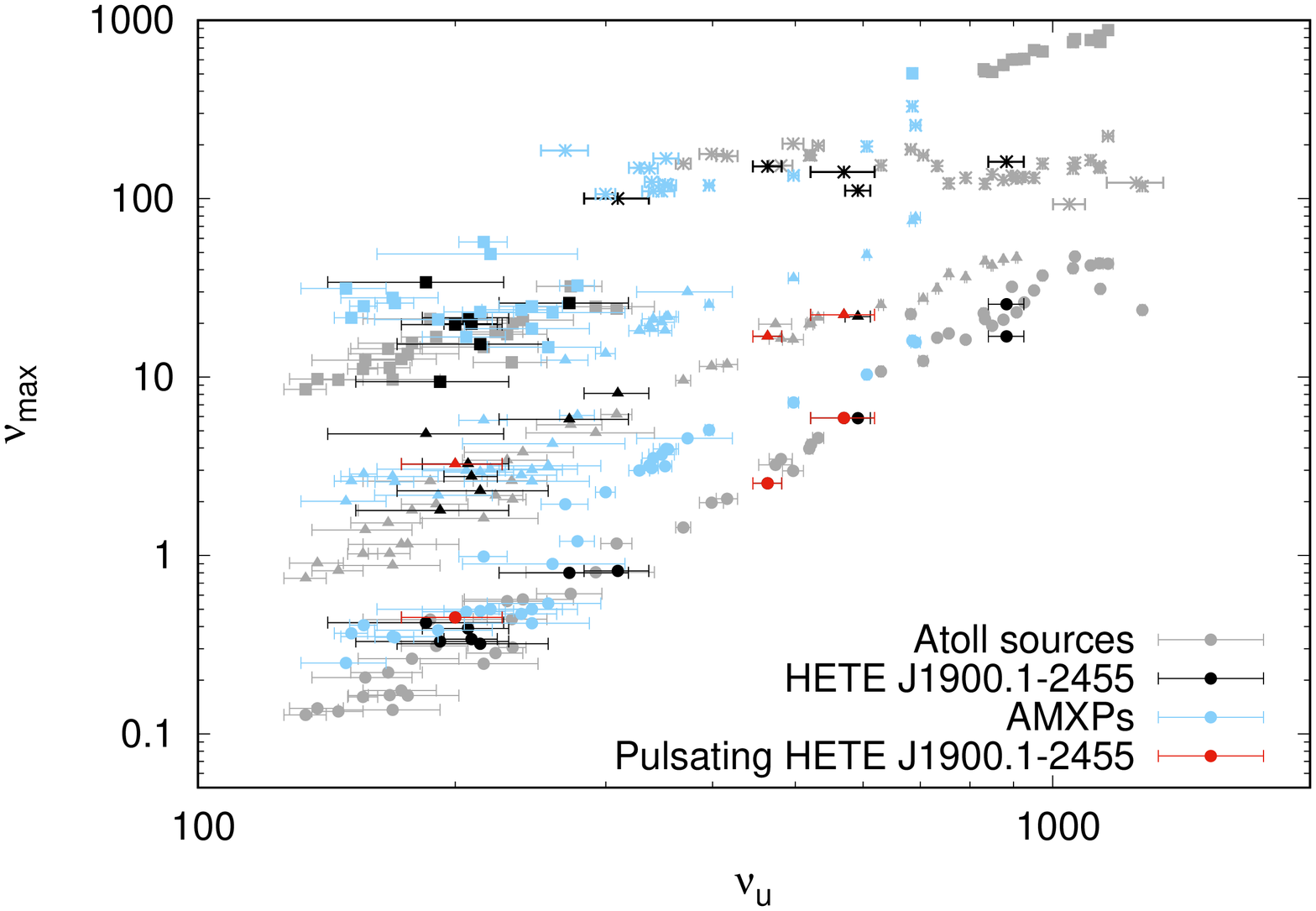}}
  \caption{Characteristic frequency of the break $L_b$ (circles), hump
    $L_h$ (triangles), hecto-Hz $L_{hHZ}$ (asterisks), and $L_{low}$
    (squares) versus the upper kHz QPO frequency. The light gray
    points are atoll sources whereas the light blue points are AMXPs
    (both data-sets are taken from \citealt{van05}). The black and red
    points are new measurements done for HETE J1900.1--2455, with data
    showing pulsations (red) and no pulsations (black). All data with
    pulsations refer to the initial 22 days of the outburst.}
  \label{fig:corr}%
\end{figure*}

We then proceed by looking at the correlations between the upper kHz
QPO ($L_u$) and the other characteristic frequencies in the power
spectra (namely $L_b$, $L_h$, $L_{hHz}$ and $L_{low}$). The results are shown in
Figure~\ref{fig:corr}, where we display only data points which have an
error on $L_u$ of less than 50 Hz, since larger errors are
not particularly constraining. This leave us with 12 groups
out of a total of 62 groups considered in this work (see Table~\ref{tab:obs}).

The data of J1900 clearly suggest that this source behaves as a
non-pulsating atoll source, even when pulsations are detected (see
Figure~\ref{fig:corr} where we show the data with pulsations in red
and without as black). Some data points are not very constraining
since the characteristic frequencies are located towards the lower end
of the observed range, where the error bars on all data points are
large and the measurements of atoll sources and AMXPs overlap
substantially. However, there are data points (some of which were
collected in the initial 22 days when the source was a persistent
pulsator) that clearly show no shift in characteristic frequencies
between J1900 and non-pulsating atoll sources.  Second, the relatively
constant frequency of the 100--200 Hz feature in
Figure~\ref{fig:corr}, along with its low Q value, suggests that
indeed we are observing the hHz QPO and we thus have no detection of a
lower kHz QPO in J1900 (see Table~\ref{tab:fit} for a summary of all
fitted components).

Finally, we looked for the parallel-track phenomenon, which is seen in
several atoll sources (see \citealt{van01} and \citealt{van06}
for a review).  When plotting the upper kHz QPO frequency versus the
average 2--16 keV luminosity we see a clear single correlation with no
parallel tracks. However, we caution that the number of measurements
is relatively small and therefore we cannot firmly conclude that the
parallel track phenomenon does not occur in J1900.

\subsection{Pulsations}

We now turn our attention to the drifting pulse episode reported by
\citet{pat12c} and observed at the time of the first flare.  The first
striking point to make is that the occurrence of the pulse phase drift
happens in coincidence with the beginning of the \textit{RXTE}
observation. The timescale for the pulsations to drift by $\approx 1$
cycle is roughly 6 ks -- i.e., about twice the length of the
observation (where the pulses are seen to drift by $\approx 0.5$
cycles). This sets an absolute upper limit to the duration of the
drift. The length of the flaring episode is ill constrained on these
timescales, since the preceding and following \textit{RXTE}
observations (when the source was not in the flare) were taken about 2
days apart (each).  Therefore it seems a remarkable (and very
unlikely) coincidence that the pulse phases drift by 0.5 cycles
exactly when the \textit{RXTE} observation takes place. Indeed if the
drift is a unique event that occurs only once during the flare, then
it should last no more than $\approx6$ ks. But the flaring duration
can be anything between 3 ks and 4 days.  If the flare has lasted for
only $\approx6$ ks, then we have a probability of less than 2\% to
observe, by coincidence, the flare at the right moment with
\textit{RXTE}.

To further investigate this phenomenon, we first look at the pulse
fractional amplitudes (sinusoidal semi-amplitude). All observations
show a relatively constant pulsed fraction (in the 2--16 keV band) at
around $2\%$ with no clear trend and/or variation over time. The
fractional amplitudes in the $\approx2-9$ keV soft band ($\approx
1.5-2\%$) are about half those in the $9-17$ keV hard energy band
($\approx 3-4\%$). When looking at the pulse phases, the hard band
pulses arrive slightly earlier than those in the soft band.  The trend
of the hard band pulse phases appears also slightly different than the
soft pulses, with the former that seem to drift more quickly in the
beginning and then reaching a maximum, while the soft pulses keep
drifting at an approximately constant rate.

\section{Discussion} 

We have used all available \textit{RXTE} data of the AMXP HETE J1900.1--2455 to
investigate its periodic and aperiodic variability properties. 
HETE J1900.1--2455 shows three peculiar phenomena, namely it has:
\begin{itemize}

\item[1.] the break and hump characteristic frequencies which scale with the upper kHz QPO frequency as non-pulsating atoll sources rather than AMXPs;
\item[2.] the highest upper kHz QPO frequency ever observed in an accreting pulsar ($\approx900$ Hz);
\item[3.] the largest pulse phase jump of any AMXP ($\approx 0.6$--$0.7$ cycles) with the pulse phases drifting by $\approx 0.5$ cycles in about 3 ks.  
\end{itemize}
It is therefore interesting to ask whether these peculiar phenomena
are related to each other and whether they might provide further
insight on the physics operating in the inner regions of the accreting
neutron star.

The shift in frequencies between power spectral components of other
AMXPs and non-pulsating atoll sources~\citep{van05} provides a
powerful tool to probe the intermittent behavior of J1900. Indeed it
is possible that these shifts are related in some way to the presence
of a dynamically important magnetosphere in AMXPs.  This would seem a
natural direction to follow when trying to build a physical
interpretation of the data.  It is important to stress that when
looking at these shifts there are exceptions to the rule. For example,
a somewhat smaller shift ($\approx1.17$) was observed in the
non-pulsating atoll source 4U 1820--30~\citep{alt05}. Another
exception is the AMXP IGR J17511--3057, where the shift factor was
varying over time and required an offset of about 2 for most
observations~\citep{kal11}. However, we stress that all AMXPs, until
now, have been observed with these shifts.  We also note that the
upper kHz QPO frequencies seen in AMXPs are shifted to lower values
with respect to other atoll sources, which is expected if the upper
kHz QPO is related to the Keplerian frequency of the inner accretion
disk. The inner disk boundary in AMXPs is indeed regulated by the
neutron star magnetic field strength that pushes it further out than
in non-magnetized neutron-stars.  Since all AMXPs show these shifts,
it is clear that a magnetosphere is the primary candidate to explain
the aperiodic variability phenomenology.

In HETE J1900.1-2455 the scaling of the characteristic frequencies
matches very well those of other non-pulsating atoll sources, both
when the source is observed pulsating persistently and when it turns
into a non-pulsating source. Therefore what is observed in J1900
argues in favor of a relatively weak magnetosphere that is unable to
shift the upper kHz frequency to lower values.  The behavior of the
other power-spectral components of J1900 also closely follows those of
other atoll sources, with the caveat that several power spectral
components appear slightly shifted in color classification. For
example, we see a $\approx883$ Hz QPO during the island state, whereas
these are usually observed in the lower left banana branch.  We also see the
hHz QPO in the extreme island state rather than the island
state.  This anomaly is, however, not unique to J1900 since some atoll
sources have shown a similar behavior (e.g., \citealt{van03, mig03})
as well as an AMXP~\citep{kal11}.

The second piece of evidence for the presence of a weak magnetosphere
comes from the very high frequency QPO observed to reach values of
$\approx 900$ Hz. If this is interpreted as a Keplerian
frequency~\citep{van00}, then this would correspond to a radius of
17--20 km for a neutron star mass of $1.4$--$2.0\msun$. If we
calculate the magnetosphere radius from the usual expression that
equates the ram pressure in the disk to the magnetic pressure of the
magnetosphere, then:
\begin{equation}
r_{{M}}= \left({\mu^4\over 2GM\dot{M}^2} \right)^{1/7}
\end{equation}
with $\mu= BR^3$ the star's magnetic moment, $M$ the neutron star's
mass and $R$ its radius. The actual location of the transition to
magnetically dominated flow is, however, more uncertain than this, and
a factor $\xi \approx 0.1-1$ is used to parametrize the uncertainties
in the physical model~\citep{gho79, psa99}.  If we set $r_{M} = 17$--$20$ km and use
$\xi=1$ for simplicity, then the inferred strength of the magnetic
field is of the order of $7$--$10\times10^{7}$ G, in line with what
was estimated by \citet{pat12c} based on the accretion torques behavior (but again we caution that a lower $\xi$ would increase $B$). 

A third piece of evidence for a weak magnetic field comes from the
behavior of the pulsations. The pulse phase drift is anomalous because
the probability of observing it by pure chance, given the spacing
between consecutive \textit{RXTE} observations of $\approx2$ days, is
extremely small, less than $2\%$. This suggests that the 3--6 ks
timescale of the drift does not correspond to the total duration of
the phenomenon which is very likely much larger than that.  The fact
that during the drift there is very little variation in pulsed
fractional amplitude strongly argues against a small hot spot
(drifting in latitude) on the neutron star surface. Indeed, although
the amplitude of the pulsations has a complex dependence on a number
of physical parameters like the hot spot size, location on the
surface, temperature differential etc. (see e.g., \citealt{pou06b}),
the small amplitude of the pulsations and the lack of amplitude
variability argue in favor of a very large hot spot so that the
fraction of neutron star surface irradiating pulsed X-rays remains
roughly constant over time. The drift in phase might indicate that the
centroid of the emitting region is quickly drifting, possibly because
of a reconfiguration of the magnetic field, so that it is the average
centroid of the hot spot that quickly moves over time rather than a
small size hot spot. If this is the correct explanation, this would
again suggest that the magnetosphere in HETE J1900.1--2455 is very
weak so that only part of the accretion flow is channeled and/or
confined by the magnetic field lines.  Finally, the origin of the
drift itself remains enigmatic.  However, whatever mechanism is
responsible for the drift, it must be related to some anomaly in the
magnetosphere since it involves the channeling of gas from the disk to
the magnetic poles.

\begin{center}
\begin{deluxetable*}{ll}
\tabletypesize{\footnotesize}
\tablewidth{0pt}
\tablecaption{{\it RXTE} observation IDs of the fit groups\label{tab:obs}}
\tablehead{ 
\colhead{Group} & \colhead{{\it RXTE} observation IDs} 
}
\startdata
1  & 91015-01-03-03, 91015-01-03-04, 91015-01-03-05, 91015-01-03-06\\
2  & 91015-01-04-04, 91015-01-04-06, 91015-01-04-07\\
3  & 91015-01-05-00\\
4  & 91015-01-06-01, 91015-01-06-02, 91015-01-07-00, 91015-01-07-01\\
5  & 91059-03-03-01, 91059-03-03-02, 91059-03-03-03, 91059-03-04-00, 91057-01-01-00, 91057-01-01-01\\
6  & 91057-01-04-01, 91057-01-04-02, 91057-01-04-03, 91057-01-04-04, 91057-01-05-00, 91057-01-05-01, 91057-01-05-02,\\
& 91057-01-05-03, 91057-01-05-04, 91057-01-06-00, 91057-01-06-01, 91057-01-06-02, 91057-01-06-03, 91057-01-06-04,\\
& 91057-01-07-00, 91057-01-07-01, 91057-01-07-02, 91057-01-07-03, 91057-01-07-04, 91057-01-08-00, 91057-01-08-01,\\
& 91057-01-08-02, 91057-01-08-03, 91057-01-08-04, 91057-01-09-00, 91057-01-09-01, 91057-01-09-03, 91057-01-10-00, \\
& 91057-01-10-01, 91057-01-10-02, 91057-01-10-03, 91057-01-11-00, 91057-01-12-00, 91432-01-01-00, 91432-01-01-01G\\
7 & 92049-01-01-00, 92049-01-02-00, 92049-01-04-00, 92049-01-05-00\\
8 & 92049-01-44-00\\
9 & 93030-01-53-00, 93030-01-54-00\\
10 & 93030-01-64-00, 93451-01-01-00, 93451-01-02-00, 93451-01-03-00, 93451-01-04-00, 93451-01-05-00, 93451-01-06-00,\\
& 93451-01-07-00, 93451-01-08-00, 93451-01-09-00, 94030-01-01-00, 94030-01-02-00\\
11 & 94030-01-20-00, 94030-01-21-00, 94030-01-22-00, 94030-01-23-00, 94030-01-24-00, 94030-01-25-00, 94030-01-26-00,\\
& 94030-01-27-00, 94030-01-28-00, 94030-01-29-00, 94030-01-30-00, 94030-01-31-00, 94030-01-32-00, 94030-01-34-00,\\
& 94030-01-36-00, 94030-01-37-00, 94030-01-38-00, 94030-01-40-00, 94030-01-41-00\\
12 & 95030-01-40-00, 95030-01-41-00, 95030-01-42-00, 95030-01-44-00, 95030-01-45-00, 96030-01-01-00, 96030-01-01-01,\\
& 96030-01-02-00, 96030-01-02-01, 96030-01-03-00, 96030-01-06-00, 96030-01-07-00, 96030-01-08-00\\

\enddata

\tablecomments{\scriptsize Subdivision of {\it RXTE} observation IDs in groups as used in this paper. The groups reported
are only those for which we have sufficient S/N to allow a secure identification of the power spectral components, out of a total of 62 groups analyzed.}

\end{deluxetable*}
\end{center}

\begin{center}
\begin{deluxetable*}{lcccccc}
\tabletypesize{\tiny}
\tablewidth{0pt}
\tablecaption{Characteristic frequencies of the multi-Lorentzian fit for HETE J1900.1--2455\label{tab:fit}} 
\tablehead{ 
\colhead{Group} & \colhead{L$_{b}$} & 
\colhead{L$_{b2}$} & \colhead{L$_{h}$} &
\colhead{L$_{\ell ow}$} & \colhead{L$_{hHz}$} &  \colhead{L$_u$} \\
  & (Hz) & 
(Hz) & (Hz) & 
(Hz) & (Hz) & 
 (Hz)}

\startdata
\hline\\
\medskip
1 & $0.451^{0.017}_{-0.014}$ & -- & $3.25\pm0.10$ & $19.7^{4.3}_{-4.1}$ &  -- & $200^{27}_{-24}$\\
\medskip
2 & $2.54^{0.22}_{-0.32}$  & -- &  $16.9^{1.0}_{-1.3}$  & -- & $152^{15}_{-12}$ & $464^{18}_{-24}$\\
\medskip
3 & $5.9\pm0.9$    &  -- & $22.3^{2.0}_{-1.8}$ & -- & $141^{39}_{-26}$ &  $570^{49}_{-52}$ \\
\medskip
4 & $5.89^{0.54}_{-0.50}$   & -- & $21.8^{1.5}_{-1.4}$  & -- & $111^{19}_{-14}$  & $592^{20}_{-30}$\\
\medskip
5 & $0.42^{0.04}_{-0.03}$   & -- & $4.80^{0.28}_{-0.26}$ & $34\pm3$ &  -- & $185^{43}_{-40}$\\
\medskip
6 & $0.337^{0.008}_{-0.007}$ & -- & $2.77\pm0.07$ & $19.9^{1.3}_{-1.4}$ & -- &  $209\pm15$\\
\medskip
7 & $0.33\pm0.04$ & -- & $1.79^{0.33}_{-0.30}$  & $9.42^{0.91}_{-0.77}$ & -- &  $192^{39}_{-29}$\\
\medskip
8 & $25.6^{1.0}_{-0.9}$ & $16.9^{9.7}_{-9.8}$ & --  & -- & $161^{29}_{-19}$ &  $883^{42}_{-36}$\\
\medskip
9 & $0.83^{0.06}_{-0.05}$ & -- & $8.1\pm0.4$  & -- & $100^{14}_{-7}$ &  $310^{26}_{-28}$\\
\medskip
10 & $0.799^{0.047}_{-0.052}$ & -- & $5.8^{0.5}_{-0.4}$  & $26^{15}_{-10}$ & -- &  $272^{47}_{-39}$\\
\medskip
11 & $0.319^{0.016}_{-0.015}$ & -- & $2.3^{0.16}_{-0.14}$  & $15.3^{2.8}_{-2.0}$ & -- &  $214^{43}_{-41}$\\
\medskip
12 & $0.389^{0.012}_{-0.014}$ & -- & $3.26^{0.22}_{-0.16}$ & $21.5^{4.2}_{-2.7}$ & -- & $207^{24}_{-21}$ \\
\hline
\enddata
\tablecomments{\scriptsize Characteristic frequencies ($\equiv 
  \nu_{\rm max}$) of the Lorentzians used to model the power spectra of HETE J1900.1--2455.
  The errors are evaluated for $\Delta\chi^{2}$ = 1.0.\\}
\end{deluxetable*}
\end{center}

\acknowledgments{AP acknowledges support from a NWO (Netherlands Organization for Scientific Research) Vidi fellowship. RW is supported by an NWO Top grant, module 1.

\end{document}